# Extending Science from Lunar Laser Ranging

A white paper submitted to
the Committee on the Planetary Science Decadal Survey (2023-2032) of
The National Academies of Sciences

Vishnu Viswanathan[1,2] *, Erwan Mazarico[1], Stephen Merkowitz[1], James G. Williams[3], Slava G. Turyshev[3], Douglas G. Currie[4], Anton I. Ermakov[5], Nicolas Rambaux[6], Agnès Fienga[6,7], Clément Courde[7], Julien Chabé[7], Jean-Marie Torre[7], Adrien Bourgoin[8], Ulrich Schreiber[9], Thomas M. Eubanks[10], Chensheng Wu[4], Daniele Dequal[11], Simone Dell'Agnello[12], Liliane Biskupek[13], Jürgen Müller[13], Sergei Kopeikin[14]

[1] *NASA Goddard Space Flight Center, 8800 Greenbelt Rd., Greenbelt, MD 20771, USA*
[2] *University of Maryland, Baltimore County, 1000 Hilltop Cir., Baltimore, MD 21250, USA*
[3] *JPL, California Institute of Technology, 4800 Oak Grove Dr., Pasadena, CA 91109, USA*
[4] *Department of Physics, University of Maryland, College Park, MD 20742, USA*
[5] *Department of Earth and Planetary Science, University of California, Berkeley, CA 94720, USA*
[6] *IMCCE, CNRS, Observatoire de Paris, PSL Université, Sorbonne Université, Paris, France*
[7] *Géoazur, CNRS, Observatoire de Côte d'Azur, Valbonne/Caussols, France*
[8] *University of Bologna, Dipartimento di Ingegneria Industriale, via Fontanelle 40, Forlì, Italy*
[9] *Institute for Astronomical and Physical Geodesy, Technische Universität München, Germany*
[10] *Space Initiatives Inc., 572 Burlington Ave. NE, Palm Bay, FL 32907, USA*
[11] *Scientific Research Unit, Italian Space Agency, Matera, Italy*
[12] *National Institute for Nuclear Physics (INFN-LNF), via E. Fermi 54, Frascati (Rome), Italy*
[13] *Institute of Geodesy, Leibniz University of Hannover, Schneiderberg 50, Hannover, Germany*
[14] *Department of Physics and Astronomy, University of Missouri, Columbia, MO 65211, USA*

Co-signers:
Dominic Dirkx (TU Delft-Netherlands), Hauke Hussmann (DLR-Germany), Alexander Stark (DLR-Germany), Vishwa Vijay Singh (IfE-Germany), Giorgio Spada (Urbino University-Italy), Daniele Melini (INGV-Italy), Sander Goossens (NASA/GSFC/UMBC-USA), Aurélien Hees (OBSPM-France), Frank Lemoine (NASA/GSFC-USA).

**\*Contact information:**
Dr. Vishnu Viswanathan
NASA Goddard Space Flight Center (UMBC)
8800 Greenbelt Rd., Greenbelt, MD 20771
Code 698 – Bldg. 34 S296A
Email: vishnu.viswanathan@nasa.gov
Tel: +1 301 614 6466

**Abstract:** The Lunar Laser Ranging (LLR) experiment has accumulated 50 years of range data of improving accuracy from ground stations to the laser retroreflector arrays (LRAs) on the lunar surface. The upcoming decade offers several opportunities to break new ground in data precision through the deployment of the next generation of single corner-cube lunar retroreflectors and active laser transponders. This is likely to expand the LLR station network. Lunar dynamical models and analysis tools have the potential to improve and fully exploit the long temporal baseline and precision allowed by millimetric LLR data. Some of the model limitations are outlined for future efforts. Differential observation techniques will help mitigate some of the primary limiting factors and reach unprecedented accuracy. **Such observations and techniques may enable the detection of several subtle signatures required to understand the dynamics of the Earth-Moon system and the deep lunar interior.** LLR model improvements would impact multi-disciplinary fields that include **lunar and planetary science**, Earth science, fundamental physics, celestial mechanics and ephemerides.

## I. Introduction:

Lunar Laser Ranging (LLR) makes precise round-trip time-of-flight measurements with short laser pulses fired from an Earth station, bounced off a retroreflector on the near-side lunar surface (see Fig. 1) and returned. These observations are affected by orbital, geophysical and relativistic phenomena and need to be interpreted using established physical models. Ranges to several lunar retroreflectors enable the monitoring of lunar orientation. An extended data span is required for the recovery of long-term signals. Model parameters can be refined to improve the fits of these reduced measurements, extend our knowledge of the Earth-Moon dynamical system, and study the Moon's interior structure. This white paper provides an overview of the science derived and state-of-the-art over the last decade of the LLR experiment, with actionable items highlighted in red. Advances in laser technology and timing systems have brought measurement precision to the present millimetric level. Model developments have been mostly incremental, with fits to the past two decades of high-precision LLR data at the 1-2 cm level. New observation techniques and future opportunities enable the expansion of the reflector and ground station networks and will be key to addressing open science questions about the complex dynamics of the Earth-Moon system.

**Open questions that impact our understanding of the evolution of the Earth-Moon system:**
- Determine the size, shape and state of the deep lunar interior from dynamical signatures.
- Improve our understanding of the lunar tidal dissipation mechanism.

## II. Science from LLR over the last decade (2013-2022) with lessons for the future:

*a. Lunar geophysics:* The success of the NASA GRAIL mission[1] has benefitted LLR-derived science. GRAIL estimates of the lunar gravity field and tidal potential Love numbers[2,3] provide strong constraints for LLR fits. As a result, new estimates of lunar interior properties were obtained, narrowing down plausible models of the lunar interior structure and properties[4]. LLR-determined tidal dissipation implies a 200 km partial melt layer at the bottom of the lunar mantle[5–8]. LLR provides the layer's dissipation $1/Q$ at two periods and upper limits at two more periods. An absorption-band relation that peaks at 3 months fits the results, but improved accuracy is desired. The presence of a solid inner core[9] is not resolved. A strong detection of the solid inner core will determine its tidal deformation properties and help explain the occurrence of an early lunar dynamo that concur with lunar thermal evolution models[10,11]. New retroreflectors and differential measurements are likely to enable detection. Knowledge of the lunar fluid core's polar oblateness from LLR allows an estimation of the radius of the lunar core-mantle boundary and the lunar free core nutation. It also helps assess the hydrostatic nature at those depths[12]. LLR analysis provides displacement Love numbers $h_2$ and $l_2$, fluid-core/solid-mantle boundary (CMB) dissipation, and moment of inertia differences. Improved estimations of these parameters help constrain the long-term evolution of the Earth-Moon system[13]. Current lunar interior parameter uncertainties are impacted by limited knowledge of the lunar density profile. We expect the Lunar Geophysical Network (LGN) New Frontiers-class mission to bridge this gap and enable further investigation of the lunar interior structure with independent and complementary lunar datasets[14].

*b. Lunar orbit, Earth's rotation and orientation:* New estimates of the secular rates of the lunar orbital elements were obtained. There is an unexplained eccentricity rate of $\sim 3 \times 10^{-12}$ per year[13]. LLR data are sensitive to Earth's latitude variations (from Chandler wobble of the Earth's rotation axis) and UT1 time determined by the rotation angle. LLR data were used for combined solutions of Earth Orientation Parameters (EOPs) with JPL's Kalman Earth Orientation Filter (KEOF)[15]. Such combined solutions show a better fit to older LLR data than the IERS C04 series[16]. Modern LLR data shows capacity to detect sub-milliarcseconds (mas) EOP inaccuracies[17]. LLR also can potentially tie the Earth-Moon reference frame to the International Celestial Reference Frame (ICRF) at comparable accuracy of Delta-Differential One-Way Ranging (Δ-DOR) based ties[18], benefiting the realization of reference systems involving other astronomical geodetic techniques.

*c. Planetary and lunar ephemerides:* Independent planetary and lunar ephemerides were generated using 50 years of LLR data[16,19,20]. Model differences between independent solutions remain, but all solutions fit the past two decades of LLR data at the 1-2 cm (rms in one-way range) level. Earlier LLR data lack observational accuracy at these levels, thus their rms residuals show greater spread. Other lunar solution differences result from choices of model parameters, weight adjustments, and fitted data span. The lunar reflector coordinates (in the lunar principal axes frame) from each solution are in agreement at a few centimeters and the geocentric lunar center position at the level of a few tens of centimeters. Differences in lunar orientation (from integrated Euler angles) are a few tens of mas. Other estimated parameters have a general agreement. Continued LLR operation and data are fundamental to the maintenance of planetary and lunar ephemerides, essential for future precise lunar mission planning and navigation capabilities.

*d. Tests of fundamental physics:* The high accuracy of LLR data enable precision tests of fundamental physics and constraints on gravity theories. New improved LLR constraints were reported on tests of violation of Lorentz symmetry parameterized under the standard-model extension (SME) field theory framework. No deviation from the theory of general relativity (GR) was reported[21,22]. LLR tests of the universality of free fall capture a combined effect from the Strong Equivalence Principle (SEP, from differences in the gravitational self-energies of the Earth and Moon) and the Weak Equivalence Principle (WEP, due to compositional differences in the two bodies). New infrared LLR data may homogenize the data distribution vs lunar phase, an influence on EP tests[23,24]. LLR analyses provide constraints on the parametrized post-Newtonian (PPN) parameters ($\beta$ and $\gamma$), geodetic precession of the lunar orbit and gravitomagnetism[25]. No deviations from GR were reported[25–27]. A new constraint on the relative temporal variation of the gravitational constant ($\dot{G}/G_0$) was also obtained[26]. Model improvements and extended data analyses will continue to serve theoretical physics by further constraining the field of possible gravity theories.

## III. Observation Techniques

*a. Current developments in operations:* Weak signals limit present operations to a select few stations on Earth due to the single-photon detection regime imposed by factors such as telescope aperture, laser power, beam divergence and throughput-loss from two-way transmission through the Earth's atmosphere, detection efficiency, etc.

In 2015, the Grasse station in France demonstrated the advantage of LLR operations at a near-infrared wavelength (1064 nm), resulting in a transmission gain ratio of up to 3.6 at low elevation angles[28]. In 2018, the Wettzell station in Germany resumed LLR operations, with millimetric-level accuracies also obtained at the near-infrared wavelength. Continued operation of such near-infrared (millimeter-level) LLR efforts would also help to homogenize the temporal and spatial distribution of the global LLR data collected by the existing LLR network. The Apache Point station in New Mexico, USA has obtained high-accuracy LLR observations for over a decade. The Matera Laser Ranging Observatory in Italy upgraded its system performance since 2017 and contributes LLR observations of centimeter-level accuracies. The Yunnan Observatory in China demonstrated LLR capability in 2018, albeit at a lower (meter-level) accuracy. LLR capabilities are being developed at the Zhuhai station in China, the Hartebeesthoek station in South Africa[29] and the Altai station in Russia[30], and considered for the Mt. Stromlo station in Canberra, Australia. New stations in the southern hemisphere offer an advantage to enhance LLR-derived science as they would enable more uniform observational coverage of the lunar declination. Operations over two hemispheres have a geometrical advantage for Earth-related parameters such as precession and nutation.

*b. Future techniques for a breakthrough in data precision:* Currently, decimeter-level discrepancies of retroreflector coordinates exist between independent lunar ephemeris solutions[18], which corresponds to a few milliarcsecond uncertainties on the lunar orientation and frame definitions. Near-instantaneous laser range tracking of multiple targets on the lunar surface from an Earth station will strengthen the resolution of the target coordinates and enable detection of weaker geophysical signals using range differences to suppress common signals including lunar orbital radius and atmospheric path delays[31,32]. High-repetition rate LLR, allowing fast switches between targets, gives a near-instantaneous measurement that can tap into these advantages. A new LLR facility at the Table Mountain Observatory in California, USA aims to use a high-power (~ 1.1 kW) continuous-wave laser to push LLR into a ~$10^3$ photons/sec regime (expected from a single 10 cm CCR), for similar benefits[33]. These techniques have potential to form a differenced-LLR observable, an optical analog to the concept of Same Beam Interferometry[34,35], with an anticipated sub-mm accuracy. Differential techniques would aid studies of the deep lunar interior.

An active lunar laser transponder maintaining line-of-sight coverage with the Earth would offer a significant gain in the signal strength ($1/R^2$ rather than $1/R^4$ dependence on the distance between the source and detector) with an ambitious potential to also expand the limited LLR network to the wider SLR network. However, transponders require power for operation, unlike passive retroreflectors. Asynchronous laser transponders with accurate clock referencing have demonstrated capabilities[36]. Their application could extend the lessons learned from LLR-derived science to interplanetary distances[37] and they can support high-precision requirements[38] for future planetary missions[39,40] and time transfer applications[41,42]. New lunar spacecraft carrying a radio ranging device (or beacon) would allow a differential measurement from multiple VLBI stations on Earth. Such techniques have been demonstrated for position location of landers[43] and orbiters[44] and have potential to also serve as navigation systems in the cislunar environment[45]. Such co-location measurements of a lander and a quasar using the VLBI network would help tie the lunar orbit to the inertial frame (ICRF), complementary to the obliquity and dynamical equinox information provided by LLR to planetary ephemeris fits.

## IV. Data accuracy and model developments

*a. How can we improve the quality of the LLR measurements?* High-quality two-way LLR data are made available to the public by personnel at Apache Point (NM, USA), Grasse (France), Matera (Italy), and Wettzell (Germany). Maintenance of data quality and monitoring of systematic errors over decades is particularly challenging. Local calibration setups/tools help to provide critical information to enhance the science derived from LLR analysis[46]. Data screening using prediction tools is the first step. Such services are offered by the Paris Observatory's Lunar Analysis Center (POLAC) or the LLR residual calculator by the Institute of Applied Astronomy of the Russian Academy of Sciences (IAA-RAS). Multi-technique observations like SLR, GPS, VLBI and gravimetry bring independent information to identify the origin of systematic local signals[47,48]. Station-specific centimeter-level biases were found through ephemeris fits to LLR data[16,19,23], few of which match the timing of local equipment changes. Lessons learned from the impact of systematic biases in SLR data on their products can be extended to LLR[49]. This will help further exploit the scientific utility of the long temporal span offered by LLR data.

*b. Standardizing of LLR data and access to analysis tools:* LLR echoes obtained over a typical observation duration (~10 mins) are compared with a predicted return time for the generation of a single "averaged" observation called a normal point, a process similar to SLR. LLR data are archived by the ILRS[50]; a curated version (in the legacy "MINI" format) is also accessible online through the POLAC (http://polac.obspm.fr). Lunar ephemerides and analysis tools are used for providing predictions and the normal point generation follows the ILRS-recommended iterative correlation algorithm. Improved algorithms are useful to detect system-specific systematic errors[51] and reduction of variance using high-precision prediction models are possibilities[52]. In a high-photon return-rate regime, normal point signatures from the LRA off-axis tilt (a major source of range accuracy for array design) can be better quantified - a method demonstrated for SLR[53]. Standardization of LLR normal points generated by independent stations will allow close monitoring of relative data accuracy, important for the quality of fit to prediction models. Such efforts require international collaboration between the observers and analysis groups that warrants both open-access data and analysis tools. Community efforts are needed to support the expanding LLR network and to maximize the multi-disciplinary science return.

*c. Present limitations imposed by the existing lunar retroreflectors:* The surface LRAs from the Apollo and Lunokhod missions continue to allow passive range operation from Earth-based stations. The Apollo corner cubes are recessed, their design was primarily motivated to minimize thermal gradients when exposed to sunlight. The Lunokhod corner cubes have silver on their three back faces that heat up in sunlight. During sunlit conditions, thermal gradients develop within both types of corner-cubes, degrading the collimation of the return beam which results in a reduced return throughput[54]. Over time, this trend stacks-up a deficit of LLR observations at the full Moon phase, resulting in a non-uniform distribution of the LLR data vs. lunar phase which impacts the science derived from LLR (e.g., reduced sensitivity to equivalence principle violation signals[27,55]). An improvement is expected from continued LLR operation in near-infrared[23]. The large size of the LRAs was initially favorable for detection at sub-meter level range accuracies, at the inception of the LLR experiment. At present, LRAs limit the range accuracy due to pulse spreading of the reflected photons, occurring when viewed at a non-normal incidence at large optical libration angles. These limitations establish the need for improved lunar retroreflector designs. A single corner-cube retroreflector (CCR) design[56] – either as a solid[57,58] or hollow/open[59,60] configuration, offers advantages in terms of their limited pulse spreading, thermal resilience and enhanced return signal strength that would significantly improve range accuracy compared to the present LRAs.

NASA has recently funded such an effort to enable the delivery of three Next-Generation Lunar Retroreflectors (NGLRs) each consisting of a single solid 10 cm CCR supporting sub-millimeter range accuracies[57]. The first deployment of these NGLRs is nominally manifest to be delivered to Mare Crisium in late 2022 by a Commercial Lunar Payload Services (CLPS) provider. The normal point accuracy achieved with LLR will then depend upon improvements in the station hardware (e.g., shorter laser pulses, improved timing electronics) and procedures (e.g., normal point algorithms). The resulting accuracy may then be tapped into for improved science results using improved models and analysis tools. Careful optical and thermal characterization of these CCRs on the ground (before launch) will ensure their longevity in the lunar environment, enable long-term range operations and support science goals[61]. An extended network of lunar retroreflectors closer to the lunar limbs and lunar poles will complement the existing geometry[56] and help monitor dynamical signatures of the interior, important to address the open science questions. Large single CCRs placed at permanently shadowed/low-temperature regions near the lunar south pole would be exposed to a much smaller lunar thermal variation, lifting thermal constraints on their size and potentially expanding the LLR station network from the throughput gained[62]. However, Earth-based LLR operations would be limited to specific time windows offering a line of sight.

*d. Current state-of-the-art models and analysis tools:* An elaborate model of the Earth-Moon system dynamics and a sophisticated data analysis tool is required for extracting the scientific utility from high-accuracy LLR data. Typical lunar model parameters include (but are not limited to) initial conditions of orbit state vectors and orientation of the Moon, geophysical parameters such as gravity fields, the lunar polar moment of inertia, moment differences, tidal displacement Love numbers, dissipation-related parameters, the shape of the lunar core-mantle boundary, and coordinates of LRAs. The state-of-the-art models are in close agreement with the last two decades of LLR data at the 1-2 cm (weighted-rms) level in one-way range. This means that model predictions are very close to the observations, but imperfections remain. For instance, a few low-degree lunar gravity field terms (3 of the 7 degree-3 gravity field spherical harmonic coefficients) can be adjusted at the 1% level from the GRAIL-derived values. This is likely to be dissipation and unmodelled effects impacting the longitudinal libration of the Moon[4,16,23]. A single time-delay model for lunar tidal dissipation in the integrator fits well with the monthly tidal periods but tidal dissipation parameters (for tidal periods at 206 days, 365 days and 1095 days) are introduced as corrections to the longitude libration. Dissipation-related terms at longer periods may be present, but more complex rheological models are needed to represent the full spectrum of the lunar dissipation mechanism[5]. Dissipation due to the liquid core (e.g., from viscous friction at the solid-liquid interface) is separated from that due to solid tides[63]. However, the presence of a lunar solid inner core (within the fluid outer core) will introduce an additional dissipative solid-liquid interface, as well as new modes analogous to those of the lunar mantle. They remain undetected with present model accuracy. Tidal energy dissipation in the Earth-Moon system also impacts secular rates of lunar orbital elements. Other model parameters relate to terrestrial and lunar orbit-related dissipation constrained by independent estimates of solid and ocean tides on Earth[13].

LLR data analysis tools follow the model recommendations of the IERS[64] for processing observations originating from Earth. For the modeling of tidal displacements of the lunar surface (sensitive to the LRAs position), the lunar degree-2 tidal displacement Love number ($h_2$) determined from LLR has small differences (at 2-$\sigma$) with that independently determined from the analysis of LRO's Lunar Orbiter Laser Altimeter data[23,65]. A model of Earth station hydrology loading is recommended when analyzing LLR observations[47]. Site displacements due to various mass loadings are accessible via the EOST/GGFC (http://loading.u-strasbg.fr) or IMLS/NASA

(http://massloading.net)[66]. Local tide measurements (e.g., using superconducting gravimeters[48]) provide a more accurate and independent constraint for LLR data analysis. Thermal expansion of the lunar regolith and the LRAs (and those onboard Lunokhod rovers) induce millimeter level signals on the range data, impacting tests of the equivalence principle with LLR[27]. A new LRA-specific thermal expansion model is available[67]. The IERS[64] recommended tropospheric delay model assumes spherical symmetry, ignoring the impact of horizontal gradients in the refractive index of the atmosphere. These gradients increase the variance on the photon arrival times, impacting range accuracy primarily at low elevation angles. Recent tropospheric delay modeling efforts that include the azimuthal asymmetry of the atmosphere show a variance reduction when analyzing SLR data[68,69]. These efforts are likely to also benefit LLR data processing.

### V. Summary of future possibilities to extend scientific return:

An extended network of passive lunar retroreflectors (single-CCRs), like the NGLRs to be deployed in 2022 (Mare Crisium), 2023 and 2024, closer to the lunar limbs and lunar poles will complement the existing geometry, support higher ranging accuracies and thus benefit LLR-derived science. Active pointing (e.g., using gimbal actuators) will help maximize the return rate and support the robotic deployment of LLR-enabled retroreflectors with actuators (recent efforts by the European Space Agency (ESA); https://exploration.esa.int/s/WmMyaoW) or active transponders. Differential measurements have potential to break new ground in terms of accuracy, giving access to the subtle dynamic signatures of the Earth-Moon system through the structure and dynamics of the deep lunar interior, enabling a deeper insight into the evolutionary history of the system. Such measurements can be challenging to achieve using the existing LLR stations. A mid-term solution would be to transition to the use of IR, due to the demonstrated advantages[28], also beneficial for SLR operations[70]. Extension of LLR stations to the southern hemisphere offers a complementary geometry. Continued support for existing stations for range measurements to both old and new retroreflectors will ensure maximum science return. Active laser transponders would offer a significant increase in signal strength, allowing an extension of LLR capability to multiple SLR ground stations, and help pave way for future missions involving interplanetary laser ranging.

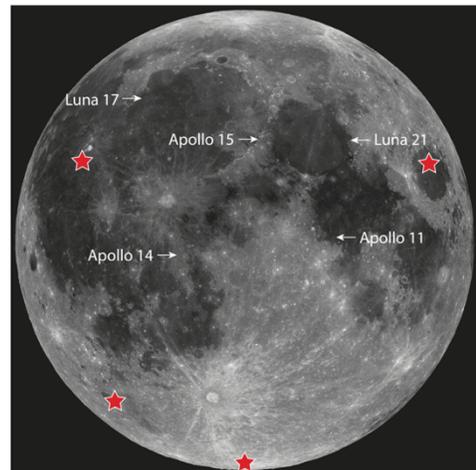

Figure 1: Existing LRA locations (labelled by mission) and possible locations (star) for future retroreflector deployment.

NASA's CLPS program and the LGN mission are expected to be the primary drivers for near-future lunar surface deployments. Developments of models and analysis tools can be accelerated through open-access efforts. Bringing together the observer and analysis community is key to ensuring data standardization, identifying systematics and improving the quality of fits (e.g., LLR workshops; http://www.issibern.ch/teams/lunarlaser). Improved LLR data and model accuracy impacts multi-disciplinary fields[71] that include lunar and planetary science, Earth science, fundamental physics, celestial mechanics and ephemerides.